\documentclass[12pt]{article}

\usepackage{graphicx}
\usepackage{amsmath}

\begin{document}

\begin{center}
{\Large {\bf
Partial Bell-state analysis with parametric down conversion
in the Wigner function formalism.}}

\vspace{1cm}

{\bf A. Casado$^1$, S. Guerra$^2$, and J. Pl\'{a}cido$^3$}.

$^1$ Departamento de F\'{\i}sica Aplicada III, Escuela Superior de
Ingenieros,

Universidad de Sevilla, 41092 Sevilla, Spain.

Electronic address: acasado@us.es

$^2$ Centro Asociado de la Universidad Nacional de Educaci\'{o}n a
Distancia de Las Palmas de Gran Canaria, Spain.

$^3$Departamento de F\'\i sica, Universidad de Las Palmas de Gran
Canaria, Spain.

\vspace{0.5cm}

(already published: Advances in Mathematical Physics, Volume 2010\\
Article ID: 501521, doi: 10.1155/2010/501521)

\vspace{1cm}
\end{center}


\vspace{1cm}
\noindent {\bf Abstract}

We apply the Wigner function formalism to partial
Bell-state analysis using polarization entanglement produced in
parametric down conversion. Two-photon statistics at a
beam-splitter are reproduced by a wavelike description with
zeropoint fluctuations of the electromagnetic field. In particular,
the fermionic behaviour of two photons in the singlet state is
explained from the invariance on the correlation properties of two
light beams going through a balanced beam-splitter.
Moreover, we show that a Bell-state measurement introduces some
fundamental noise at the idle channels of the analyzers.
As a consequence,
the consideration of more independent sets of vacuum modes
entering the crystal appears as a need for a complete
Bell-state analysis.

\vspace{1cm}

\newpage

\section{Introduction}
The theory of parametric down-conversion (PDC) in the Wigner
formalism, along with the theory of detection, was treated
in a series of papers \cite{pdc2, pdc4, pdc5, pdc6}.
The formalism was applied to experiments exhibiting
relevant aspects of quantum mechanics, such as entanglement,
nonlocality, and other nonclassical features of light.
In contrast to the usual Hilbert space formulation, where the
corpuscular nature of light is stressed, the Wigner formalism
resembles classical optics. More specifically, this alternative approach
takes into account the coupling between the zeropoint field (ZPF) and the laser
beam entering the nonlinear crystal.
Moreover, the propagation of the light fields through the different optical
devices is completely classical.
A formal bridge between classical nonlinear optics and the quantum theory
within the Wigner approach involves two elements without classical
counterpart, such as the ZPF itself, entering into the crystal and the rest
of optical devices, and the detection process, in which those vacuum
fluctuations are substracted, giving rise to the typically quantum results.

The development of quantum information in recent years, alongside with the
important role of parametric down conversion for experimental
schemes, has motivated the application of the Wigner approach to
some relevant contexts, up to now almost exclusively linked to the Hilbert domain.
Examples of these are quantum cryptography \cite{gisin}, dense coding \cite{Bennett}
and teleportation \cite{telep}.
This research program seeks to apply an alternative explanation for that kind
of phenomena, an explanation nevertheless consistent with quantum theory, and
therefore with the usual Hilbert space formulation.

Recently, the Wigner formalism has been applied to experiments on quantum
cryptography based on the Ekert's protocol \cite{ekert}, also including
the presence of eavesdropping in the case of projective measurements.
There, it was shown that the Heisenberg uncertainty principle, a key aspect in
secure quantum cryptography, is related to the change of the correlation
properties of light fields, after going through the optical devices in Alice's
and Bob's setups.
These correlation properties are affected by vacuum modes activated in the
crystal, which give rise to quantum entanglement.
Furthermore, the action of Eve introduces some noise, that also turns out
to be fundamental to reproduce the quantum results \cite{pdc7}.

Bell-state measurements constitute another key aspect in the field of quantum
information, posing a relevant problem in quantum dense coding and
teleportation schemes.
In this context,
entangled photon pairs produced in parametric down conversion have also
been used in the last decades for experiments on partial Bell-state measurement
\cite{mattle}, in which entanglement involves only one degree of freedom, and
complete Bell-state measurement, in which hyperentanglement (entanglement between
two or more degrees of freedom) takes part \cite{kwiatweinfurter}.

The paper is organized as follows:
In section $2$, we introduce the general description of the four Bell-states
within the Wigner framework. This description involves the manipulation of only
one beam. In this same section, we also study two-photon statistics at a balanced
beam-splitter.
In section $3$ we study an experiment on partial Bell-state analysis \cite{mattle}.
Finally, in section $4$, we discuss the results, making a comparison with the
Hilbert space description.

\section{Two-photon statistics at a beam-splitter in the Wigner approach}
We shall start this section by reviewing the basic concepts in the Hilbert
framework \cite{Ekert}. The four Bell states (entangled in polarization)
produced in the process of PDC are:

\begin{equation}
|\Psi^{\pm}\rangle =\frac{1}{\sqrt{2}}[|H\rangle_1|V\rangle_2\pm
|V\rangle_1|H\rangle_2],
\label{hyper1}
\end{equation}
\begin{equation}
|\Phi^{\pm}\rangle=\frac{1}{\sqrt{2}}[|H\rangle_1|H\rangle_2\pm
|V\rangle_1|V\rangle_2],
\label{hyper2}
\end{equation}
where $H$ ($V$) represents linear horizontal (vertical) polarization.
Let us suppose the two beams are recombined at a balanced beam
splitter ($BS$). If $|a\rangle$ and $|b\rangle$ represent the input
modes of the $BS$, the possible spatial states are:

\begin{equation}
|\psi_A \rangle=
\frac{1}{\sqrt{2}}[|a\rangle_1|b\rangle_2- |b\rangle_1|a\rangle_2],
\,\,\,\,;\,\,\,\,
|\psi_S \rangle=
\frac{1}{\sqrt{2}}[|a\rangle_1|b\rangle_2+ |b\rangle_1|a\rangle_2],
\label{hyper8}
\end{equation}
where $A$ ($S$) denotes antisymmetric (symmetric). Due to the fact
that the particles carrying the information are photons, the total
state must obey the bosonic symmetry, so that the total two photon
states are $|\Psi^+\rangle|\psi_S \rangle$, $|\Psi^-\rangle|\psi_A
\rangle$, $|\Phi^+\rangle|\psi_S \rangle$, $|\Phi^-\rangle|\psi_S
\rangle$.
Assuming that the $BS$ does not influence the internal state (polarization),
the two-photon state can only be changed in the spatial part, via the Hadamard
transformation, as $\hat{H}|a\rangle=(1/\sqrt{2})(|a\rangle+|b\rangle)$ and
$\hat{H}|b\rangle=(1/\sqrt{2})(|a\rangle-|b\rangle)$. Owing to
$\hat{H}|\psi_A\rangle=|\psi_A \rangle$, only in this case the two
photons emerge at the different outputs from the BS. In the other
three cases, the two photons emerge together in one of the two
outputs ports \cite{loudon}.

Let us now go to the Wigner formalism.
The Wigner transformation stablishes a correspondence between a
field operator acting on a vector in the Hilbert space and a
(complex) amplitude of the field.
In the case of zeropoint field, these amplitudes follow a particular
stochastic distribution, given by the Wigner function of the vacuum.

Quantum predictions corresponding to the state $|\Psi^+\rangle$ are
reproduced in the Wigner framework by considering the following two correlated
beams outgoing the crystal \cite{pdc4}:

\begin{equation}
{\bf F}_{1}^{(+)}({\bf r}, t)=F_s^{(+)}({\bf r},
t; \{\alpha_{{\bf k}_1, H}; \alpha^*_{{\bf k}_2, V}\}){\bf
i}+F_{p}^{(+)}({\bf r}, t;
\{\alpha_{{\bf k}_1, V}; \alpha^*_{{\bf k}_2, H}\}){\bf j}
\label{hyper0},
\end{equation}
\begin{equation}
{\bf F}_{2}^{(+)}({\bf r}, t)=F_{q}^{(+)}({\bf r},
t; \{\alpha_{{\bf k}_2, H}; \alpha^*_{{\bf k}_1, V}\}){\bf
i'}+F_{r}^{(+)}({\bf r}, t;
\{\alpha_{{\bf k}_2, V}; \alpha^*_{{\bf k}_1, H}\}){\bf j'},
\label{hyper3}
\end{equation}
where ${\bf i}$ and ${\bf i'}$ (${\bf j}$ and ${\bf j'}$) are unit
vectors representing horizontal (vertical) linear polarization at
beams ``$1$" and ``$2$", and
$\{\alpha_{{\vec k}_i, V};\alpha_{{\vec k}_i, H}\}$\,($i=1, 2$) represent
four sets of relevant zeropoint amplitudes entering the crystal. The four
set of modes $\{{\vec k}_{i, \lambda}\}$\,($i=1, 2$;  $\lambda\equiv H, V$)
are ``activated" and coupled with the laser beam inside the nonlinear medium.

As said before,
amplitudes $\{\alpha_{{\vec k}, \lambda}\}$ follow a distribution
given by the Wigner function of the vacuum field \cite{pdc2}:

\begin{equation}
W_{\rm ZPF}(\{\alpha\})=
{\prod_{{\bf [k]}, \lambda}}\frac{2}{\pi}
{\rm e}^{-2|\alpha_{{\bf k}, \lambda}|^2}.
\label{eq_w9}
\end{equation}
If $A({\bf r}, t; \{\alpha\})$ and $B({\bf r'}, t'; \{\alpha\})$ are
two complex amplitudes, the correlation between them is given by:

\begin{equation}
\langle AB \rangle=\int W_{\rm ZPF}(\{\alpha\})A({\bf r}, t; \{\alpha\})B({\bf r'}, t';
\{\alpha\})d\{\alpha\}.
\label{corr}
\end{equation}

In expressions (\ref{hyper0}) and (\ref{hyper3}),
the only non vanishing correlations are those involving the combinations
$r \longleftrightarrow s$ and $p \longleftrightarrow q$.
These correlations are directly related to the way in which the
vacuum components are distributed inside the total field amplitudes.

The four Bell-states can be generated by manipulating only
one beam, and this is related to the possibility of sending two bits
of classical information via the manipulation of only one particle
\cite{Bennett}.
In the Wigner framework, the effect of a linear optical device on a
beam accounts for a change on the distribution of zeropoint amplitudes
inside the field components. Therefore, correlation properties are also
changed.
In ref. \cite{pdc7} we performed the same analysis on the four Bell
states, this time considering a modification of the two beams, initially
departing from the description of $|\Psi^+\rangle$.
In this paper, in order to keep consistency with the essence of dense coding,
we will consider that the optical devices modify only one of the beams,
while the other beam remains unchanged from its generation at Bob's station.

Let us now focus on the experimental setup in figure \ref{Fig.1}.
The transformations at Bob's station are performed by a polarization
rotator and a wave retarder.
For instance, if we place a polarization rotator acting on beam ``$1$",
the plane of polarization of ${\bf F}_{1}^{(+)}$ will be rotated by an
angle $\beta$.
We can compute the field components behind the rotator in the following
way:

\[
{\bf F'}_{1}^{(+)}({\bf r}, t)
=\begin{pmatrix}
  {\rm cos}\beta & -{\rm sin}\beta \\
  {\rm sin}\beta & {\rm cos}\beta
\end{pmatrix}
\begin{pmatrix}
F_s^{(+)}({\bf r}, t) \\ F_{p}^{(+)}({\bf r}, t)
\end{pmatrix}
\]
\begin{equation}
=
\begin{pmatrix}
 F_s^{(+)}({\bf r}, t) {\rm cos}\beta-F_{p}^{(+)}({\bf r}, t){\rm sin}\beta \\
 F_s^{(+)}({\bf r}, t) {\rm sin}\beta+F_{p}^{(+)} ({\bf r}, t){\rm cos}\beta
\end{pmatrix}.
\label{hyper4}
\end{equation}

Now, let the wave retarder introduce a phase shift $\kappa$ between the
horizontal and vertical field components of beam ``$1$". Taking into account
the action of both optical devices, the expressions of the beams are:

\[
{\bf F''}_{1}^{(+)}({\bf r}, t)=[F_s^{(+)}({\bf r}, t){\rm cos}\beta-F_{p}^{(+)}({\bf r}, t){\rm sin}\beta]
{\bf i}
\]
\begin{equation}
+{\rm e}^{i\kappa}[F_s^{(+)}({\bf r}, t){\rm
sin}\beta+F_{p}^{(+)}({\bf r}, t){\rm cos}\beta]
{\bf j},
\label{hipo1}
\end{equation}
\begin{equation}
{\bf F}_{2}^{(+)}=F_{q}^{(+)}({\bf r}, t){\bf i'}+F_{r}^{(+)}({\bf r}, t){\bf
j'}.
\label{hyper5}
\end{equation}

The combination $\beta=0$, $\kappa=0$ gives (\ref{hyper0}) and (\ref{hyper3}),
corresponding to the state $|\Psi^+\rangle$.
On the other hand, for $\beta=0$ and $\kappa=\pi$ we obtain the
description of $|\Psi^-\rangle$:

\begin{equation}
{\bf F}_{1}^{(+)}({\bf r}, t)=F_s^{(+)}({\bf r}, t){\bf i}-F_{p}^{(+)}({\bf r}, t)
{\bf j}\,\,\,\,;\,\,\,\,{\bf F}_{2}^{(+)}({\bf r}, t)=F_{q}^{(+)}({\bf r}, t){\bf i'}+F_{r}^{(+)}({\bf r}, t){\bf
j'}.
\label{psimenos}
\end{equation}
In both cases the horizontal component of one beam is correlated with
the vertical component of the other, the only difference being the minus sign
that appears in ${\bf F''}_{1}^{(+)}$ in the case of $|\Psi^-\rangle$.

Finally, the case $\beta=-\pi/2$ and $\kappa=\pi$ corresponds to the description
of $|\Phi^+\rangle$

\begin{equation}
{\bf F}_{1}^{(+)}({\bf r}, t)=F_{p}^{(+)}({\bf r}, t)
{\bf i}+F_s^{(+)}({\bf r}, t)
{\bf j}\,\,\,\,;\,\,\,\,
{\bf F}_{2}^{(+)}({\bf r}, t)=F_{q}^{(+)}({\bf r}, t){\bf i'}+F_{r}^{(+)}({\bf r}, t){\bf
j'},
\label{fimas}
\end{equation}
and $\beta=-\pi/2$, $\kappa=0$ corresponds to the description of
$|\Phi^-\rangle$:

\begin{equation}
{\bf F}_{1}^{(+)}({\bf r}, t)=F_{p}^{(+)}({\bf r}, t)
{\bf i}-F_s^{(+)}({\bf r}, t)
{\bf j}\,\,\,\,;\,\,\,\,
{\bf F}_{2}^{(+)}({\bf r}, t)=F_{q}^{(+)}({\bf r}, t){\bf i'}+F_{r}^{(+)}({\bf r}, t){\bf
j'}.
\label{simenos}
\end{equation}
In these two cases we observe that the horizontal (vertical)
component of one beam is correlated with the horizontal (vertical)
component of the other, the difference being the minus sign that
appears in ${\bf F}_{1}^{(+)}$ in the case of $|\Phi^-\rangle$.

This description of the four Bell-states is equivalent to the one in \cite{pdc7}.
Nevertheless, as we already pointed out, in this case we have modified only one
of the two beams.
The net effect of the polarization rotator and the wave retarder is
similar to the one of a half-wave plate and a quarter-wave plate, used in
\cite{mattle}.

The general expressions (\ref{hipo1}) and (\ref{hyper5}), where the values of
$\beta$ and $\kappa$ are undetermined, correspond, in the Hilbert space, to a
superposition of the base states $|\Psi^+\rangle$, $|\Psi^-\rangle$, $|\Phi^+\rangle$
and $|\Phi^-\rangle$.

We will now study the action of a balanced beam-splitter on the
correlation properties of the light beams. For the sake of clarity,
we suposse an identical distance separating the source from the
$BS$'s, so the contribution of the phase shift in equation
$(16)$ of \cite{pdc4} can be from here on ignored in our calculations.

Because there is one beam at each input port, it is not necessary
to consider the vacuum field at the beam-splitter \cite{mandel}.
This time the BS does not introduce any additional noise to the one
provided by the zeropoint field entering the crystal.
The beams are represented by eqs. (\ref{hipo1}) and (\ref{hyper5}),
${\bf r}={\bf r}_{BS}$, ${\bf r}_{BS}$ being the position of the
beam-splitter where the two beams are recombined.

For the light beams at the outgoing channels we have:
\begin{eqnarray}
{\bf F}_{1, out}^{(+)}({\bf r}_{BS}, t)
&=&
F_{1H}^{(+)}({\bf r}_{BS}, t){\bf
i}+F_{1V}^{(+)}({\bf r}_{BS}, t){\bf j},
\\
\label{hyper5_0}
{\bf F}_{2, out}^{(+)}({\bf
r}_{BS}, t)
&=&
F_{2H}^{(+)}({\bf r}_{BS}, t){\bf i'}+F_{2V}^{(+)}({\bf
r}_{BS}, t){\bf j'},
\label{hyper5a}
\end{eqnarray}
where
\begin{eqnarray}
F_{1H}^{(+)}({\bf r}_{BS}, t)
&=&
\frac{1}{\sqrt{2}}[iF_{q}^{(+)}({\bf r}_{BS}, t)+F_s^{(+)}({\bf r}_{BS}, t){\rm cos}\beta-F_{p}^{(+)}({\bf r}_{BS}, t){\rm
sin}\beta],
\nonumber\\
\\
\label{hyper5c}
F^{(+)}_{1V}({\bf r}_{BS}, t)
&=&
\frac{1}{\sqrt{2}}\left[
iF_{r}^{(+)}({\bf r}_{BS}, t)+{\rm e}^{i\kappa}[F_s^{(+)}({\bf
r}_{BS}, t){\rm sin}\beta+F_{p}^{(+)}({\bf r}_{BS}, t){\rm cos}\beta]
\right],
\nonumber\\
\\
\label{hyper5d}
F_{2H}^{(+)}({\bf r}_{BS}, t)
&=&
\frac{1}{\sqrt{2}}[F_{q}^{(+)}({\bf r}_{BS}, t)+iF_s^{(+)}({\bf r}_{BS}, t){\rm cos}\beta-iF_{p}^{(+)}({\bf r}_{BS}, t){\rm
sin}\beta],
\nonumber\\
\\
\label{hyper5b}
F_{2V}^{(+)}({\bf r}_{BS}, t)
&=&
\frac{1}{\sqrt{2}}\left[F_{r}^{(+)}({\bf r}_{BS}, t)+i{\rm e}^{i\kappa}[F_s^{(+)}({\bf r}_{BS}, t){\rm sin}\beta+F_{p}^{(+)}({\bf r}_{BS}, t){\rm
cos}\beta]
\right].
\nonumber\\
\label{hyper5cc}
\end{eqnarray}

We now calculate the cross correlations between the components of ${\bf F}_{1,
out}^{(+)}({\bf r}_{BS}, t)$ and those of ${\bf F}_{2,out}^{(+)}({\bf r}_{BS}, t)$:

\begin{enumerate}
\item The correlation between the field components corresponding to the
same polarization at the outgoing channels vanishes, independently of
the values of $\kappa$ and $\beta$.
This is due to the fact that the contribution to the correlation of the
transmitted components of the field is cancelled by the contribution of
the reflected components:

\begin{equation}
\langle F_{1H}^{(+)}({\bf r}_{BS}, t)F_{2H}^{(+)}({\bf r}_{BS}, t) \rangle=-\frac{1}{2}{\rm sin}\beta
\langle F_{p}^{(+)}({\bf r}_{BS}, t)F_{q}^{(+)}({\bf r}_{BS}, t)\rangle [1+i^2]=0,
\label{hyper5e}
\end{equation}
\begin{equation}
\langle F_{1V}^{(+)}({\bf r}_{BS}, t)F_{2V}^{(+)}({\bf r}_{BS}, t) \rangle=\frac{1}{2}{\rm e}^{i\kappa}{\rm sin}\beta
\langle F_{r}^{(+)}({\bf r}_{BS}, t)F_{s}^{(+)}({\bf r}_{BS}, t)\rangle [1+i^2]=0.
\label{hyper5f}
\end{equation}

\item For the correlation between the field components corresponding to different
polarization and different outgoing channels, $\langle F_{2H}^{(+)}({\bf r}_{BS},
t)F_{1V}^{(+)}({\bf r}_{BS}, t)\rangle$ and $\langle F_{2V}^{(+)}({\bf r}_{BS},
t)F_{1H}^{(+)}({\bf r}_{BS}, t)\rangle$, we have:

\begin{eqnarray}
&
\langle F_{1H}^{(+)}({\bf r}_{BS}, t)F_{2V}^{(+)} ({\bf r}_{BS}, t)\rangle
=
-\langle F_{1V}^{(+)}({\bf r}_{BS}, t)F_{2H}^{(+)}({\bf r}_{BS}, t) \rangle
\nonumber\\
&
=
\frac{{\rm cos}\beta}{2} (\langle F_{s}^{(+)}({\bf r}_{BS},
t)F_{r}^{(+)}({\bf r}_{BS}, t)\rangle+i^2{\rm e}^{i\kappa}
\langle F_{q}^{(+)}({\bf r}_{BS}, t)F_{p}^{(+)}({\bf r}_{BS}, t)\rangle).
\nonumber\\ 
\label{hyper5g}
\end{eqnarray}

As there is no path difference before the beam-splitter
between ${\bf F}_{1, out}^{(+)}$ and ${\bf F}_{2, out}^{(+)}$, the
cross correlations $\langle F_{s}^{(+)}({\bf r}, t)F_{r}^{(+)}({\bf
r'}, t')\rangle$ and $\langle F_{q}^{(+)}({\bf r'},
t')F_{p}^{(+)}({\bf r}, t)\rangle$, computed at the same position
${\bf r}_{BS}$ and time $t$, have the same value.
With all this:

\begin{eqnarray}
&
\langle F_{1H}^{(+)}({\bf r}_{BS}, t)F_{2V}^{(+)} ({\bf r}_{BS}, t)\rangle
=
-\langle F_{1V}^{(+)}({\bf r}_{BS}, t)F_{2H}^{(+)}({\bf r}_{BS}, t) \rangle
\nonumber\\
&
=
\frac{{\rm cos}\beta}{2}
\langle F_{s}^{(+)}({\bf r}_{BS}, t)F_{r}^{(+)}({\bf r}_{BS}, t)\rangle
(1+i^2{\rm e}^{i\kappa}).
\label{hyper5g1}
\end{eqnarray}

We can see that if $\beta=\pi/2$, i.e., the states $|\Phi^\pm\rangle$,
the latter correlations vanish.
On the other hand, in the case of $|\Psi^+\rangle$ ($\kappa=0$ and $\beta=0$)
these correlations are also null.
Finally, only when $\beta=0$ and $\kappa=\pi$, i.e., the state $|\Psi^-\rangle$,
these correlations are different from zero:

\begin{eqnarray}
&
\langle F_{1H}^{(+)}({\bf r}_{BS}, t)F_{2V}^{(+)} ({\bf r}_{BS}, t)\rangle
=
-\langle F_{1V}^{(+)}({\bf r}_{BS}, t)F_{2H}^{(+)} ({\bf r}_{BS}, t)\rangle
\nonumber\\
&
=
\langle F_{s}^{(+)}({\bf r}_{BS}, t)F_{r}^{(+)}({\bf r}_{BS},
t)\rangle.
\label{hyper5i}
\end{eqnarray}

\item To conclude, we compute the correlations between the two field
components of different polarization, corresponding to the same
outgoing beam:

\begin{eqnarray}
&
\langle F_{1H}^{(+)}({\bf r}_{BS}, t)F_{1V}^{(+)} ({\bf r}_{BS}, t)\rangle
=
\langle F_{2H}^{(+)}({\bf r}_{BS}, t)F_{2V}^{(+)}({\bf r}_{BS}, t)\rangle
\nonumber\\
&
=
\frac{i{\rm cos}\beta}{2}
(\langle F_{s}^{(+)}({\bf r}_{BS},t)F_{r}^{(+)}({\bf r}_{BS}, t)\rangle+{\rm e}^{i\kappa}\langle
F_{q}^{(+)}({\bf r}_{BS}, t)F_{p}^{(+)}({\bf r}_{BS}, t)\rangle)
\nonumber\\
&
=
\frac{i{\rm cos}\beta}{2}\langle F_{s}^{(+)}({\bf r}_{BS},
t)F_{r}^{(+)}({\bf r}_{BS}, t)\rangle (1+{\rm e}^{i\kappa}).
\label{hyper5j}
\end{eqnarray}
It can be easily seen that the correlations above are different from
zero only in for the state $|\Psi^+\rangle$ ($\kappa=0$ and $\beta=0$).
In this situation:

\begin{eqnarray}
&
\langle F_{1H}^{(+)}({\bf r}_{BS}, t)F_{1V}^{(+)} ({\bf r}_{BS}, t)\rangle
=
\langle F_{2H}^{(+)}({\bf r}_{BS}, t)F_{2V}^{(+)}({\bf r}_{BS}, t)\rangle
\nonumber\\
&
=
i\langle F_{s}^{(+)}({\bf r}_{BS}, t)F_{r}^{(+)}({\bf r}_{BS},t)\rangle.
\label{hyper5k}
\end{eqnarray}

\end{enumerate}

\section{Partial Bell-state analysis in the Wigner approach}

\begin{figure}
\begin{center}
\includegraphics[height=5cm]{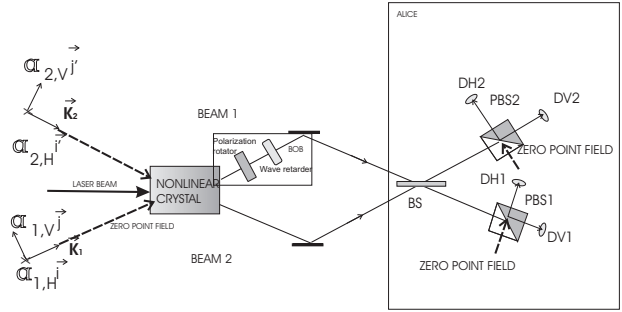}
\end{center}
\caption{
Setup for quantum dense coding. The polarization rotator and the wave
retarder used by Bob allow a transmission to Alice of any of the
Bell states. Alice's station consists on a balanced beam-splitter,
two polaryzing beam-splitters, and detectors $DH1$, $DV1$, $DH2$ and
$DV2$.}
\label{Fig.1}
\end{figure}

Let us again consider the situation in figure \ref{Fig.1}.
The nonlinear medium is an element of a quantum-optical experimental
scheme for dense coding \cite{mattle}.
The two beams (\ref{hyper0}) and (\ref{hyper3}) are correlated through
the zeropoint field entering the crystal, which is ``amplified" via the
activation of the four set of vacuum modes $\{{\vec k}_{i,\lambda}\}$\,
($i=1, 2$;  $\lambda\equiv H, V$).
The beam ``$1$" can be modified at Bob's station, who can activate the
polarization rotator and/or the wave retarder.
Manipulation of beam ``$1$" allows for the possibility of distributing
the vacuum amplitudes in four different ways, so that the correlation
properties of beams ``$1$" and ``$2$" can be modified and used for information
encoding.
In our framework, the possibility of sending two bits of classical information
via the manipulation of one particle is explained through the change on the
correlation properties of two beams when one of them is modified at Bob's station.
Such correlations have their origin in the crystal, where the zeropoint modes are
coupled with the laser field, and the information is carried by the amplified vacuum
fluctuations \cite{louisell}.

The two beams are recombined at Alice's Bell-state analyser by means
of a balanced beam-splitter ($BS$), and the horizontal and vertical
polarization components of each outgoing beams are separated at the polarizing
beam-splitters $PBS1$ and $PBS2$.
Finally, all coincidence detection probabilities can be measured with
the detectors $DH1$, $DH2$, $DV1$ and $DV2$.

We focus now on the fields at the detectors.
We will suppose there is the same distance between the $BS$ and any of
the detectors, and again the phase factor corresponding to the propagation
of the different amplitudes is irrelevant.
Owing to the fact that each polaryzing beam-splitter reflects (transmits)
the horizontal (vertical) polarization, the electric field at the detector
($DH1$, $DV1$, $DH2$, $DV2$) is the superposition of ($iF_{1H}^{(+)}$,
$F_{1V}^{(+)}$, $iF_{2H}^{(+)}$, $F^{(+)}_{2V}$) and a vacuum field
amplitude which has been (reflected, transmitted, reflected,
transmitted) at the other idle channel of the corresponding PBS.
These vacuum amplitudes have no correlation with the signals, nor they
have with each other.
If ${\bf F}^{(+)}_{ZPF, Alice1}$ (${\bf F}^{(+)}_{ZPF,Alice2}$) is the
vacuum field entering $PBS1$ ($PBS2$), for the field amplitudes at the detectors
we have:

\begin{equation}
{\bf F}_{DH1}^{(+)}({\bf r}_{DH1}, t_1)= iF_{1H}^{(+)}({\bf r}_{DH1},
t_1){\bf i}+[{\bf F}^{(+)}_{ZPF, Alice1}({\bf r}_{DH1}, t_1)\cdot
{\bf j}]{\bf j},
\label{hyper5c1}
\end{equation}

\begin{equation}
{\bf F}^{(+)}_{DV1}({\bf r}_{DV1}, t'_1)=F_{1V}^{(+)}({\bf r}_{DV1},
t'_1){\bf j}+i[{\bf F}^{(+)}_{ZPF, Alice1}({\bf r}_{DV1}, t'_1)\cdot
{\bf i}]{\bf i},
\label{hyper5d1}
\end{equation}

\begin{equation}
{\bf F}_{DH2}^{(+)}({\bf r}_{DH2}, t_2)= iF_{2H}^{(+)}({\bf r}_{DH2},
t_2){\bf i'}+[{\bf F}^{(+)}_{ZPF, Alice2}({\bf r}_{DH2}, t_2)\cdot
{\bf j'}]{\bf j'},
\label{hyper5b1}
\end{equation}

\begin{equation}
{\bf F}^{(+)}_{DV2}({\bf r}_{DV2}, t'_2)=F_{2V}^{(+)}({\bf r}_{DV2},
t'_2){\bf j'}+i[{\bf F}^{(+)}_{ZPF, Alice2}({\bf r}_{DV2}, t'_2)\cdot
{\bf i'}]{\bf i'},
\label{hyper5c1}
\end{equation}
where $F_{1H}^{(+)}$, $F_{1V}^{(+)}$, $F_{2H}^{(+)}$, $F^{(+)}_{2V}$
are given by (\ref{hyper5c}), (\ref{hyper5d}), (\ref{hyper5b}),
and (\ref{hyper5cc}) respectively.

To calculate joint detection probabilities we use (see equation $(28)$
of \cite{pdc4}):

\begin{equation}
P_{AB}({\bf r}, t; {\bf r}', t') \propto
\sum_{\lambda}\sum_{\lambda'} |\langle F_{\lambda}^{(+)}(\phi_A; {\bf
r},t) F_{\lambda'}^{(+)}(\phi_B; {\bf r}' ,t')\rangle|^2,
\label{F}
\end{equation}
where $\lambda$ and $\lambda'$ are polarization indices, and $\phi_A$
and $\phi_B$ represent controllable parameters of the experimental
setup.

For instance, let us show the calculation of $P_{DH1, DH2}$. For
simplicity we focus on the ideal situation with $t_1=t_2$, and discard
the dependence on position and time. We have:

\begin{eqnarray}
&
P_{DH1, DH2}
\propto
|\langle F_{1H}^{(+)}F_{2H}^{(+)} \rangle|^2+
|\langle F_{1H}^{(+)}[{\bf F}^{(+)}_{ZPF, Alice2}\cdot
{\bf j'}]
\rangle|^2
+
\nonumber\\
&
|\langle [{\bf F}^{(+)}_{ZPF, Alice1}\cdot
{\bf j}]F_{2H}^{(+)}\rangle|^2+ |\langle [{\bf F}^{(+)}_{ZPF, Alice1}\cdot
{\bf j}][{\bf F}^{(+)}_{ZPF, Alice2}\cdot
{\bf j'}]\rangle|^2=|\langle F_{1H}^{(+)}F_{2H}^{(+)} \rangle|^2,
\nonumber\\
\label{Fa}
\end{eqnarray}
where we have taken into account that the ZPF inputs at the PBS's are
uncorrelated with the signals and with each other. With (\ref{hyper5e}),
we finally obtain:

\begin{equation}
P_{DH1, DH2}=0.
\label{Fb}
\end{equation}

The rest of the probabilities can be obtained similarly.
By using (\ref{hyper5f}), (\ref{hyper5g1}) and (\ref{hyper5j}), we find:

\begin{equation}
P_{DV1, DV2}=0.
\label{Fc}
\end{equation}

\begin{equation}
\frac{P_{DH1, DV2}}{k_{DH1, DV2}}=
\frac{P_{DV1, DH2}}{k_{DV1, DH2}}=
\frac{{\rm cos}^2\beta}{2}[1+{\rm
cos}(\kappa+\pi)]|\langle F_s^{(+)}F_r^{(+)}\rangle|^2,
\label{Fe}
\end{equation}

\begin{equation}
\frac{P_{DH1, DV1}}{k_{DH1, DV1}}=
\frac{P_{DH2, DV2}}{k_{DH2, DV2}}=
\frac{{\rm cos}^2\beta}{2}(1+{\rm
cos}\kappa)|\langle F_s^{(+)}F_r^{(+)}\rangle|^2,
\label{Fd}
\end{equation}
where $k_{DH1, DV2}$, $k_{DV1, DH2}$, $k_{DH1, DV1}$ and $k_{DH2,
DV2}$ are constants related to the effective efficiency of the
detection processes.

From (\ref{Fb}), (\ref{Fc}),  (\ref{Fe}) and  (\ref{Fd}) we conclude:

\begin{itemize}
\item Recording a coincidence of $DH1$ and $DH2$
($DV1$ and $DV2$) is not possible, for whatever
$\kappa$ and $\beta$.

\item Recording a coincidence of $DH1$ and $DV2$
($DV1$ and $DH2$) is possible only in the case $(\beta=0,
\kappa=\pi)$, i.e. the state $|\Psi^-\rangle$.

\item Recording a coincidence of $DH1$ and $DV1$ ($DH2$ and $DV2$)
is possible only for $(\beta=0, \kappa=0)$, i.e., the state
$|\Psi^+\rangle$.

\item When $\beta=\pi/2$, i.e. the states $|\Phi^\pm\rangle$,
all coincidence probabilities vanish. Hence, these Bell-states
cannot be distinguished.

\end{itemize}

\section{Discussion}
We have applied the Wigner approach to study two-photon statistics at a balanced
beam splitter. We have also treated an experimental setup for partial Bell-state
analysis.
The Wigner formalism allows for an interpretation of these experiments
in terms of waves, but, however, the whole formalism lies inside the quantum domain,
the zeropoint field being an alternative to the role of vacuum fluctuations
in the Hilbert space.

As we already pointed out, once in the Wigner framework,
the typical quantum results appear precisely as a consequence of the introduction
of the zeropoint field. This vacuum field enters in the crystal and also in the
rest of optical devices. Finally, it is substracted in the detection process.
Quantum correlations can be then explained solely in terms of the propagation
of those vacuum amplitudes through the experimental setup, and their
subsequent substraction at the detectors.

At the beginning of section $2$ we presented the fundamental ideas on
two-photon statistics at a beam-splitter in the Hilbert space formalism,
which account for the usual corpuscular description of light.
In the Wigner framework, a clear counterpart to that description is found
when, in order to preserve the bosonic character of the photons,
the spatial part of the quantum state is forced to remain
antisymmetric. This happens only for $|\Psi^-\rangle$, and, in this
case, the pair of photons behave as fermions at the beam-splitter,
emerging at different output ports. For the other three Bell states,
in which both the polarization and spatial parts of the two-photon
state are symmetric, both particles emerge together at the same
output port of the $BS$.

For the Wigner representation, the corpuscular aspect of light appears
as just an interplay of (Maxwell) waves, including a zeropoint vacuum
field.
The action of the beam-splitter must be treated as in the classical
framework: a part of each entering beam is transmitted, and the other part
is reflected, without any change in their polarization properties.
From (\ref{hyper5g1}), it can be seen that the correlation between the
field components corresponding to different polarization and different
outgoing channels vanishes for $\beta=\pi/2$ (states $|\Phi^{\pm}\rangle$),
and also for $\kappa=0$ (state $|\Psi^{+}\rangle$).
In this last case, the factor $i^2$ indicates that the contribution to the
correlation of the transmitted components is cancelled by the contribution
of the reflected components.
Nevertheless, in the case of $|\Psi^{-}\rangle$ there is a constructive
superposition of the two terms in (\ref{hyper5g1}).
In other words, the net effect of the phase shift $\kappa=\pi$ and the
beam-splitter in the field amplitudes is to leave correlations unchanged.
This is equivalent, in the Hilbert space, to the fact that $|\psi_A\rangle$
(see eq. (\ref{hyper8})) is an eigenstate of the Hadamard operator.

At this point it is worth to stop again at the question:
how can we explain a typical particle behaviour (the bosonic nature of
photons in the Hilbert space description), from a wavelike description,
just with the inclusion of vacumm fluctuations of the electromagnetic field?
By inspection of (\ref{hipo1}) and (\ref{hyper5}), it can easily be seen that,
only in the case of $|\Psi^-\rangle$, the exchange $1 \longleftrightarrow 2$
implies a sign flip in the correlation properties of the beams (they remain
equal in the other three cases).
The bosonic nature of photons is completely represented by the correlation
properties that characterize each of the four Bell-states.
If now Bob activates just the wave retarder,
this gives rise to a change, not only in the internal (polarization)
state, but also in the spatial part of the quantum state, in order to
keep the requirement of boson symmetry.
This ``double" effect is explained in the Wigner formalism by taking into
account the way in which the zeropoint field is coupled inside the crystal
\cite{pdc4}.

The Bell-state measurement performed at Alice's station only identifies
the states $|\Psi^-\rangle$ and $|\Psi^+\rangle$. We complete our picture
here by pointing out the analyzer introduces some fundamental noise at the
idle channels of the polaryzing beam splitters.
These zeropoint fluctuations, although they will (again) be finally substracted
at the detectors, and (again) hold no correlation with the signals,
turn out to be the fundamental ingredient of the Wigner approach to this experimental
setups.
A quick look at figure \ref{Fig.1} shows that there are four relevant sets of vacuum
modes entering the crystal, in which the quantum information is ``carried", and another
two sets of vacuum modes entering the PBS's, containing only zeropoint field.
This image constrasts to the usual ``qubit language", where ``particles carry the
information" \cite{Ekert}.

Recently, the problem of performing complete Bell-state measurements has been solved
by considering a higher number of degrees of freedom (hyperentanglement).
Hyperentanglement is also a convenient resource for some other
recent and important applications in the field of quantum computing.
An example of these is one-way quantum computation using clusters states \cite{clusters}.
The use of Hilbert spaces of higher dimension is related, within the Wigner formalism,
to the inclusion of more sets of vacuum modes entering the crystal.
With increasing number of vacumm inputs, the possibility for extracting more
information from the zeropoint field also increases.
For instante, the momentum-polarization hyperentanglement needs, in the Wigner
framework, the consideration of $8$ sets of relevant modes of the zeropoint field
entering the crystal \cite{bienal2009}.

\section{Acknowledgements}

The authors thank Professor E. Santos for helpful suggestions and
comments on the work. They also thank R. Risco and D. Rodr\'{\i}guez for
their comments and careful reading of the manuscript.
A. Casado acknowledge support from the Spanish MCI Project No. FIS2008-05596.

\end{document}